\documentclass{jetpl}
\twocolumn

\lat

\title{Dimerization in honeycomb Na$_2$RuO$_3$ under pressure: a DFT study}

\rtitle{Dimerization in honeycomb Na$_2$RuO$_3$ under pressure: a DFT study}

\sodtitle{Dimerization in honeycomb Na$_2$RuO$_3$ under pressure: a DFT study}

\author{D.\,D.\,Gazizova$^{*+}$\/\thanks{e-mail: darya.05.02@mail.ru},
A.\,V.\,Ushakov$^{+}$, S.\,V.\,Streltsov$^{*+}$}

\rauthor{D.\,D.\,Gazizova, A.\,V.\,Ushakov, S.\,V.\,Streltsov}

\sodauthor{Gazizova, Ushakov, Streltsov}

\address{$^*$Ural Federal University, Mira Str. 19, 620002 Ekaterinburg, Russia\\~\\
$^+$Institute of Metal Physics, Russian Academy of Science, S. Kovalevskaya Str. 18, 620041 Ekaterinburg, Russia}

\dates{06 March 2018}{*}

\abstract{The structural properties of Na$_2$RuO$_3$ under pressure are studied using density functional theory within the generalized gradient approximation (GGA). We found that one may expect a structural transition at $\sim$ 3 GPa. This structure at the high-pressure phase is exactly the same as the low-temperature structure of Li$_2$RuO$_3$ (at ambient pressure) and is characterized by the $P2_1/m$ space group. Ru ions form dimers in this phase and one may expect strong modification of the electronic and magnetic properties in Na$_2$RuO$_3$ at pressure higher than 3 GPa.}

\PACS{74.50.+r, 74.80.Fp}

\begin{document}

\maketitle

\section{Introduction}
Compounds with the honeycomb lattice are under intensive study in last two decades, since many of them show intriguing and sometimes rather unexpected physical properties. There are materials demonstrating the spin gap behaviour as, e.g., (Na, Li)$_3$Cu$_2$SbO$_6$~\cite{Xu-05,Miura-06,Morimoto-06}, $\alpha-$MoCl$_3$~\cite{McGuire-17} and systems having different types of the long range magnetic order in the low-temperature (LT) region: zigzag antiferromagnet (AFM) as, e.g., Na$_2$Co$_2$TeO$_6$~ \cite{Lefransois-16,Bera-17}, Neel AFM (Li$_2$MnO$_3$~\cite{Lee2012,Korotin2015} and possibly SrRu$_2$O$_6$~\cite{Streltsov2015a}), and even ferromagnetic honeycomb planes, as in Ni$_3$TeO$_6$~\cite{Zivkovic2010}. Moreover, they can host an unusual spin-liquid ground state as was proposed by Kitaev~\cite{Kitaev2006}. While initial candidate, Na$_2$IrO$_3$~\cite{Jackeli2009}, does not seem to be a physical realization of the Kitaev model, another system, $\alpha-$RuCl$_3$, is under a close investigation now~\cite{banerjee-16,zheng-17,banerjee-2017}.

Another very important class of materials having honeycomb lattice are the systems, where this lattice turns out to be dimerized at the LT phase. One of the examples is Li$_2$RuO$_3$. This is a layered material. In the low-temperature (LT) phase two out of six Ru--Ru bonds in a hexagon dimerize, which results in formation of the spin gap~\cite{Miura2007}. With increase of temperature Li$_2$RuO$_3$ exhibits an unusual phase transition at $T_d \sim 540 K$, which was initially thought as a transition from a dimerized to uniform structure~\cite{Miura-09}. However, more careful study using X-ray pair distribution function analysis shows that dimers as rigid units survive even at $T>T_d$, while in the average this system can be described as undimerized, uniform (C2/m space group)~\cite{Kimber-14}. Thus, this transition at $T_d$ was described as a transition to the so-called valence-bond liquid phase, where these Ru--Ru dimers start to flow over the honeycomb lattice~\cite{park}.

Surprisingly, similar compound Na$_2$RuO$_3$ order magnetically below $T_N \sim 30$ K (AFM zigzag) without any sizeable structural distortions in honeycomb lattice \cite{Na,Gapontsev-17}. In high-temperature phase Na$_2$RuO$_3$ crystallizes in the same C2/m space group as Li$_2$RuO$_3$, but the volume of the unit cell in Na$_2$RuO$_3$ is $\sim 5\%$ larger. Thus, one might expect that under the pressure Na$_2$RuO$_3$ may start to dimerize and exhibits all unusual properties as its sister compound Li$_2$RuO$_3$. 

In the present work we study a possible formation of the dimerized crystal structure in Na$_2$RuO$_3$ under pressure using first-principle calculations using density functional theory (DFT) in the generalized gradient approximation (GGA). We found, that at $\sim 3$ GPa dimerized phase becomes lower in energy than uniform structure without any Ru--Ru dimers. Thus, we predict the phase transition to dimerized structure in Na$_2$RuO$_3$ at $\sim 3$ GPa, which may result in strong modification of the electronic and magnetic properties of Na$_2$RuO$_3$ related to destroy of the long range magnetic order in the LT phase and formation of the spin gap. 
\begin{figure}[t!]
\centering
\includegraphics[width=0.5\textwidth]{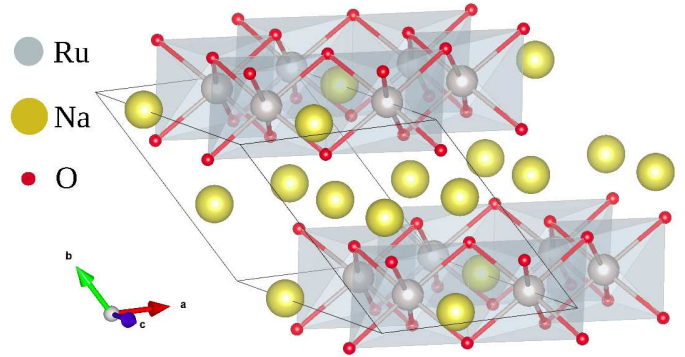}
\caption{{\bf Fig. 1.} The crystal structure of Na$_{2}$RuO$_{3}$. Ru ions are shown as grey, Na as yellow and O as red balls.}
\label{na2ruo3}
\end{figure}

\section{Calculation details}
The band structure calculations of Na$_{2}$RuO$_{3}$ were mostly performed using the Vienna Ab-initio Simulation Package (VASP)~\cite{VASP,V}. We utilized the projector augmented-wave (PAW) method~\cite{PAW} with the Perdew-Burke Ernzerhof (PBE) GGA functional~\cite{PBE}. 
The energy cutoff was chosen to be $E_{cutoff} \sim 500$ eV and the 4$\times$4$\times$4 Monkhorst-Pack grid of k-points was used in the calculations. Analysis of the electronic structure was performed within the TB-LMTO-ASA code, which is based on the linearized muffin-tin orbital method~\cite{Andersen-84}.

In this work we considered two crystal structures: uniform and dimerized ones. The uniform structure was taken from known experimental data~\cite{Na}. The dimerized phase of Na$_{2}$RuO$_{3}$ was obtained from corresponding Li$_{2}$RuO$_{3}$ structure~\cite{Miura2007}, where the Li ions were substituted by the Na ions and volume of the unit cell was rescaled accordingly.  For the crystal structure optimization we used conjugate gradient algorithm. 

\section{Crystal structure}
The crystal structure of  Na$_{2}$RuO$_{3}$ consists of the magnetic honeycomb RuO$_2$ layers in the $ab$ plane, which are separated by nonmagnetic sodiums (see Fig.~\ref{na2ruo3}). Other Na ions are in the hexagons formed by edge sharing RuO$_{6}$ octahedra.  According to the chemical formula Ru ions have a valency $4+$ with four electrons on the Ru $t_{2g}$ orbitals, which results in $S=1$. 

At ambient pressure Na$_{2}$RuO$_{3}$ (non-dimerized lattice) crystallizes in the monoclinic structure (space group $C$2/$m$). The lattice constants and atomic coordinates were determined in Ref.~\cite{Na}: $a$=5.3456(1)~\AA, $b$=9.2552(18)~\AA, $c$=5.5504(1)~\AA, $\beta$=108.74(3)$^\circ$. The unit cell volume is 260.046 \AA$^{3}$. There are two slightly different Ru-Ru bonds in the $ab$ plane: $a_{1}$ = 3.084~\AA, $a_{2}$ = 3.089~\AA, so that $r = a_{2}/a_{1} \sim 1$.

\begin{figure}[t!]
\centering
\includegraphics[width=0.5\textwidth]{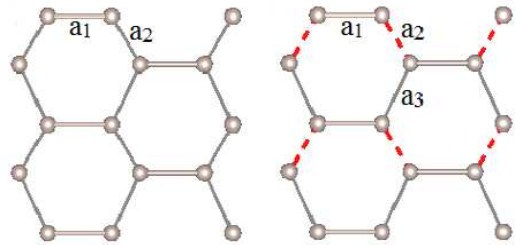}
\caption{{\bf Fig. 2.} The Ru-Ru bonds in the uniform (left part) and dimerized (right part) phases. The short Ru-Ru bonds are marked by the red dashed lines.}
\label{ru}
\end{figure}

We used the LT Li$_{2}$RuO$_{3}$ structure, which is described by the $P$2$_1$/$m$ space group~\cite{Miura2007}, to create dimerized Na$_{2}$RuO$_{3}$ structure. There are two nearly the same Ru--Ru bonds in the hexagon $a_{1}$=3.045~\AA and $a_{3}$=3.049~\AA and one short $a_{2}$=2.568~\AA, so that the ratio between longest and shortest bonds is $r \sim 1.2$ in Li$_{2}$RuO$_{3}$ at the LT phase. Shortest Ru-Ru bonds form the herringbone structure. Schematic representations of both uniform and dimerized phases is shown in Fig.~\ref{ru}.


\section{Calculation results}
In this study we carried out the DFT calculations for non-dimerized (experimental) and dimerized (artificial) structures for several unit cell volumes. At each point the shape of the unit cell was fixed and positions of the Na, Ru, and O ions were optimized. Analysis of the calculation results shows that both structures survive in chosen volume range and do not transform to any other structure.  

\begin{figure}[h!]
\centering
\includegraphics[width=0.5\textwidth]{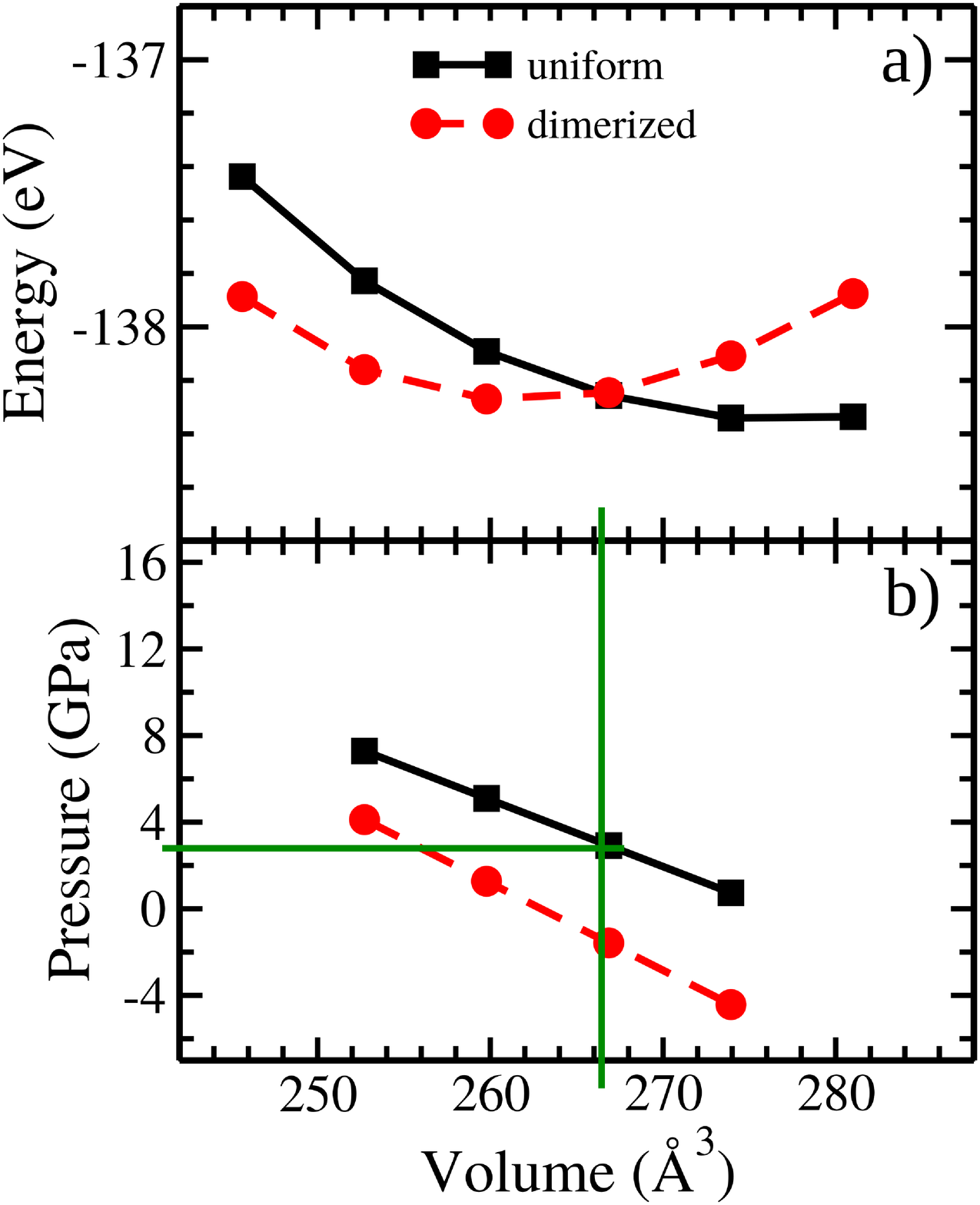}
\caption{\label{graf}{\bf Fig. 3.} (a) Total energy dependence of Na$_2$RuO$_3$ of two structures (with and without structural Ru-Ru dimers) for different volumes. (b) Pressure dependence on the volume for these two structures.}
\end{figure}

The volume dependence of total energies, $E$, is presented in Fig.~\ref{graf}(a). One can see that the uniform structure of Na$_2$RuO$_3$, which is experimentally observed at ambient pressure, does have the lowest total energy and this structure corresponds to the global minimum of $E(V)$ in the GGA. However, the situation changes with decrease of the volume and the dimerized structure turns out to be the lowest in energy for $V <  266$~\AA$^{3}$. Intristingly, the ratio between longest and shortest Ru--Ru bonds in dimerized case is $r = a_{2}/a_{1} \sim 1.2$, very close to what we have in Li$_2$RuO$_3$.

There were debates whether the structural structural transition in Li$_2$RuO$_3$ is of first or second order~\cite{Miura2007,Segura-2016}, but recent differential thermal analysis demonstrates a heat anomaly at transition temperature, which is in favour of first order transition. In order to estimate critical pressure and volume jump for similar transition in Na$_2$RuO$_3$ we first interpolated $E(V)$ (to have a smooth function) and then recalculated the pressure dependence on volume as $P = - \partial E / \partial V$, which is shown in Fig.~\ref{graf}(b).  Applying external pressure to undimerized phase of Na$_2$RuO$_3$ we reduce volume of the unit cell down to critical value $V_{undim}$, when the total energies of uniform and dimerized phases are the same. This point defines a critical pressure for transition, $P_c \sim 3$ GPa, see  Fig.~\ref{graf}(b). Further compression drives the system to the dimerized phase. Corresponding jump in the unit cell volume at the transition is $\delta V = V_{undim}-V_{dim}\sim$10 \AA$^3$. Similar transition in Li$_2$RuO$_3$ (but caused by temperature) gives $\delta V \sim 8 $\AA$^3$~\cite{Miura2007}.

One may expect that in Na$_2$RuO$_3$ there also (as in Li$_2$RuO$_3$) can develop a spin gap in the dimerized phase, so that one might expect a decrease of magnetic susceptibility at $P_c$, and the electronic structure of Na$_2$RuO$_3$ can also be very different. One cannot estimate reliably the value of the spin gap using the GGA calculations, but it is rather straightforward to study the electronic structure of this phase. We chose for this different calculation scheme - so called tight-binding linearized muffin-tin orbitals method, as realized in the TB-LMTO-ASA code~\cite{Andersen-84}. In contrast to VASP used for the volume optimization, in TB-LMTO-ASA one can easily change a local coordinate system for any ion. This helps in analysis of the electronic structure.

\begin{figure}[t!]
\centering
\includegraphics[width=0.5\textwidth]{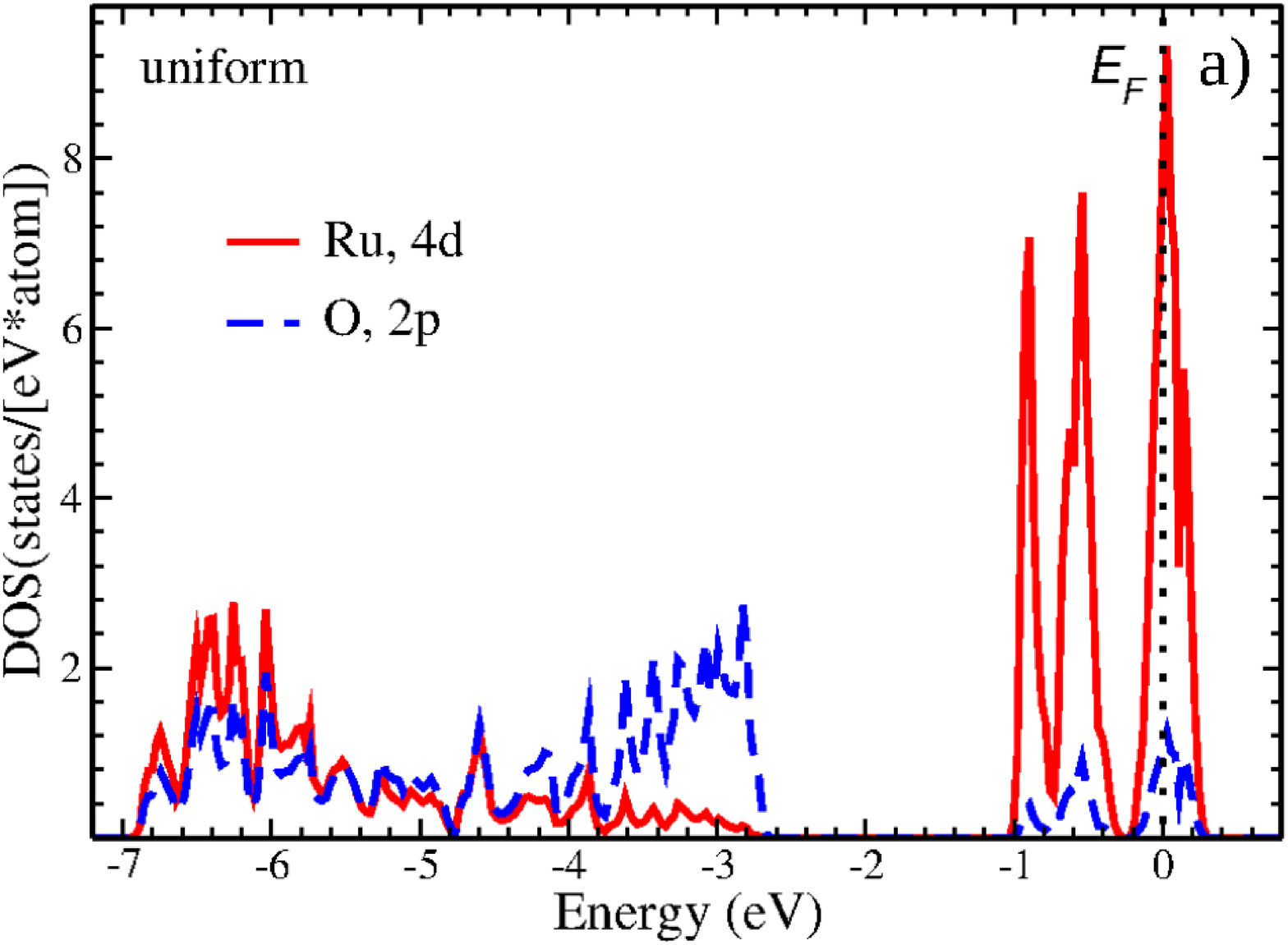}\\
\includegraphics[width=0.5\textwidth]{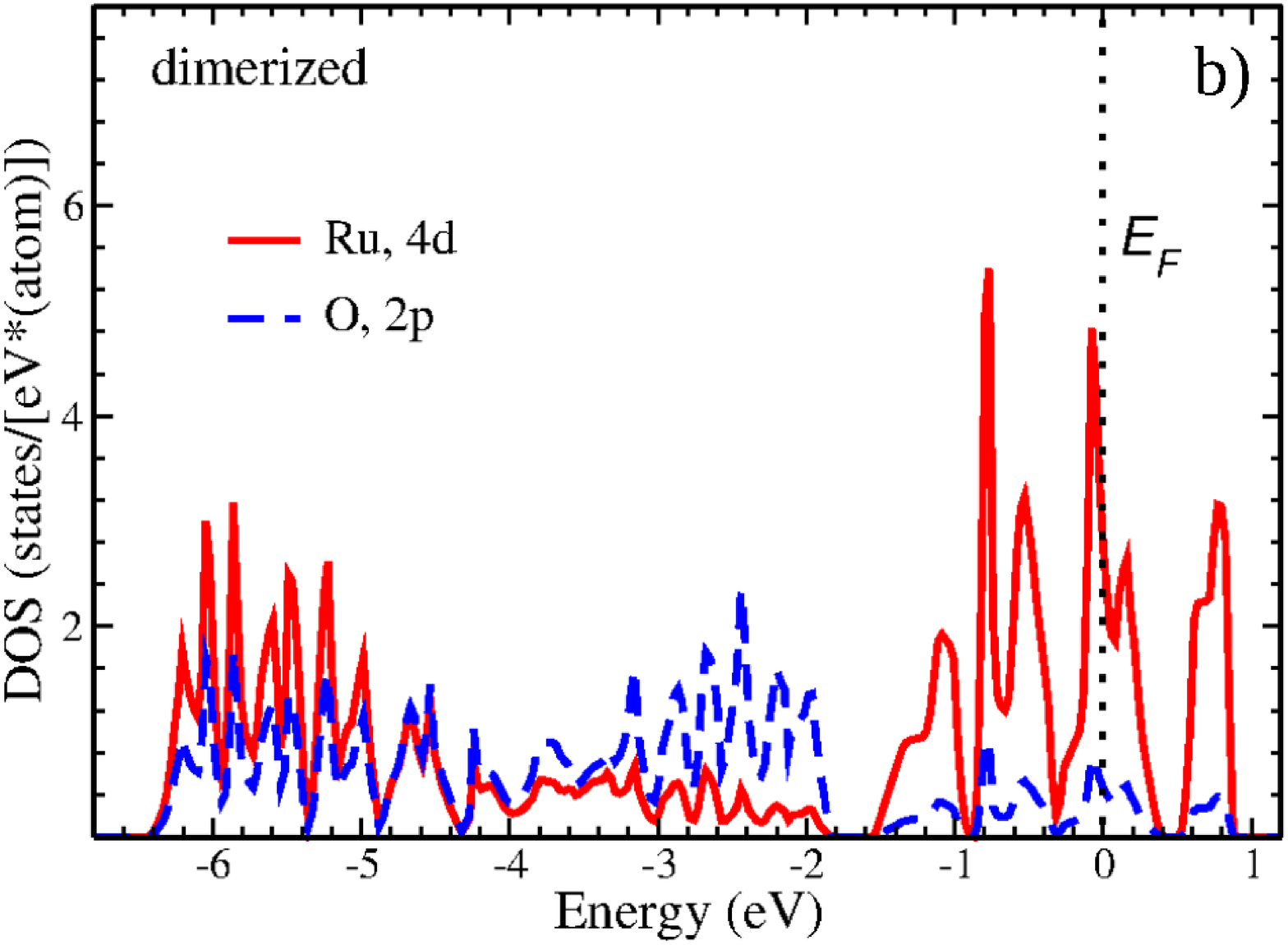}\\
\includegraphics[width=0.5\textwidth]{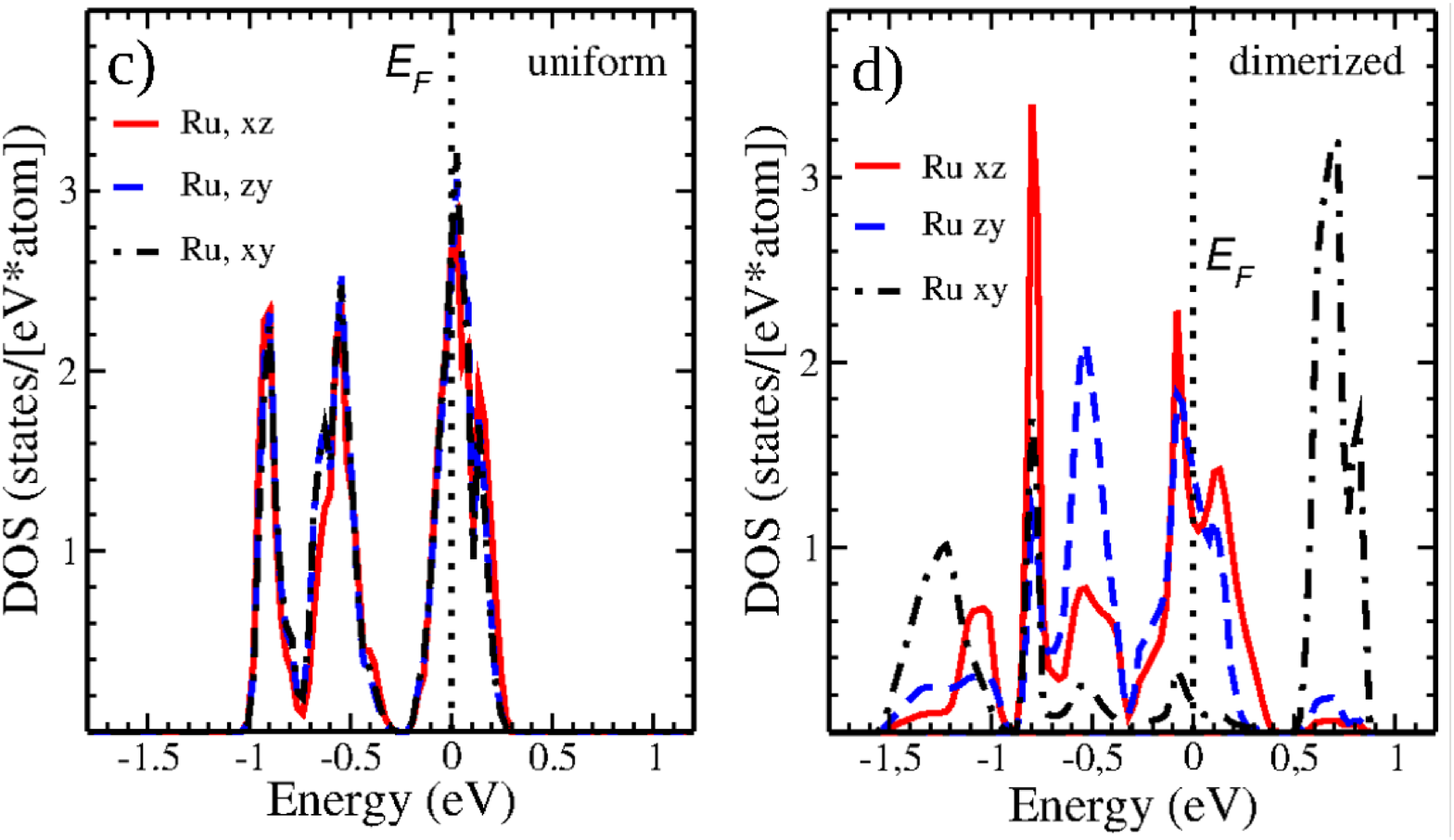}
\caption{\label{dos} {\bf Fig. 4.} Partial densities of states (DOS) of the Ru-$4d$ and O-$2p$ shells and partial DOS, corresponding to the Ru $t_{2g}$ orbitals for the uniform and dimerized Na$_2$RuO$_3$ structures at the transition volume. The local coordinate system where $x$,$y$, and $z$ axes point to the ligands in RuO$_6$ octahedron was chosen. The Fermi energy is in zero. }
\end{figure}

The partial density of state (DOS) plots for the Ru-$4d$ and O-$2p$ states for Na$_{2}$RuO$_{3}$ in the dimerized phase at the transition volume ($\sim 267$~\AA$^{3}$ on the unit cell), as calculated in the TB-LMTO-ASA, are presented in Fig.~\ref{dos}(b). One may see that in the GGA approximation Na$_{2}$RuO$_{3}$ is a metal, which is due to the absence of strong electronic correlations in this method~\cite{Martin-book}. The O-$2p$ bands are in a region from -6.5 eV to -1.8 eV, whereas the Ru-$t_{2g}$ states are located mostly in the energy interval from -1.5 eV to 1 eV. 

The electronic structure in the vicinity of the Fermi level in the dimerized phase is shown in Fig.~\ref{dos}(c). We used the local coordinate system with coordinate axes pointing to the oxygen ions. One may notice a strong bonding-antibonding splitting for part of the Ru $t_{2g}$ states. These are the $xy$ orbitals, which look to each other in the common edge geometry~\cite{Kimber-14,Streltsov-2016,Streltsov-UFN}. This large bonding-antibonding splitting is due to a direct overlap between the $xy$ orbitals on two different sites forming Ru-Ru dimers. The splittings of the $zx/zy$ states are much smaller. One may compare corresponding splittings in dimerized phases of Na$_2$RuO$_3$ and Li$_2$RuO$_3$. The bonding-antibonding splitting between the $xy$ orbitals in Na$_2$RuO$_3$ is  $\sim 2$ eV, i.e. on $\sim 0.6$ eV smaller than in Li$_{2}$RuO$_{3}$~\cite{Kimber-14}. The dimers are mainly stabilized by the formation of these bonding orbitals and thus one may see that even at pressure Ru-Ru dimers are much less rigid and stable in Na$_2$RuO$_3$ than in Li$_{2}$RuO$_{3}$. This also suggests that one may expect a smaller spin gap at low temperatures in the high-pressure phase of Na$_2$RuO$_3$ (smaller than in  Li$_{2}$RuO$_{3}$).

In contrast to the dimerized case the width of the t$_{2g}$ band in the uniform Na$_2$RuO$_3$ at the transition volume is much smaller (in $\sim 2$ times, see Fig.~\ref{dos}(c) and (d)) because of the absence of bonding-antibonding splitting. Also the O-$2p$ and Ru-$4d$ states are separated by the energy gap about $1.6$ eV.



In the end one needs to comment on possible importance of the strong electronic correlations in Na$_2$RuO$_3$. We have seen that the GGA overestimates equilibrium volume of the uniform phase on $\sim$6\% (experimental unit cell volume is 260 \AA$^{3}$ whereas the minimum of the total energy in the GGA  corresponds to $\sim 276$~\AA$^{3}$; this is rather typical that the local density approximation (LDA) underestimates, while the GGA overestimates equilibrium volume~\cite{Martin-book}). An account of the strong electronic correlations, which are typical for the transition metal ions in insulating oxides, may, possibly, improve agreement between a theory and an experiment. However, the simplest LDA+$U$-like methods~\cite{Anisimov-1991} will not help in this situation. This method tends to stabilize a solution with local spin moments forming a long range magnetic order and breaks molecular-orbitals, which are a driving force of a dimerization. As a result LDA+$U$ would always favour an uniform, non-dimerized structure. More appropriate in this situation would be cluster LDA+DMFT calculations~\cite{Biroli-2002}, which take into account on equal footing both a formation of molecular-orbitals and strong Coulomb correlations. However, presently it is impossible to calculate forces and optimize the crystal structure within the cluster LDA+DMFT approach for real materials.

\section{Conclusion}
To sum up, in the present paper we predict a structural phase transition to dimerized phase in Na$_2$RuO$_3$ at pressure $\sim$ 3 GPa. One may expect that this transition will be accompanied by strong changes in the magnetic and electronic properties, related to the formation of the spin gap and strong bonding-antibonding splitting. These changes might be detected by the susceptibility measurements and by the X-ray and optical spectroscopy. Interestingly similar transition (and at similar pressures $\sim$ 3 GPa) to the dimerized phase was recently observed in $\alpha-$Li$_2$IrO$_3$~\cite{Hermann-2018}, which also (as Na$_2$RuO$_3$) has undimerized lattice at normal conditions.

\section*{Acknowledgment}
The authors would like to thank I. Mazin, R. Valenti, and D. Khomskii for numerous discussions of physical properties of 213 ruthenates. Present work was supported  
by the project of the Ural branch of RAS 18-10-2-3, 
by the FASO through research program ``spin'' AAAA-A18-118020290104-2,
by Russian ministry of education and science via contract 02.A03.21.0006 and 
by the Russian foundation for basic research (RFBR) via grant RFBR 16-02-00451.


\begin{thebibliography}{99}



\bibitem{Xu-05} J. Xu, A. Assoud, N. Soheilnia , S. Derakhshan, H. L. Cuthbert, J.E. Greedan, M.H. Whangbo, and H. Kleinke, Inorg. Chem. \textbf{44}, 5042 (2005).

\bibitem{Miura-06} Y. Miura, R. Hirai, Y. Kobayashi, and M. Sato, J. Phys. Soc. Jpn. \textbf{75}, 084707 (2006).

\bibitem{Morimoto-06} K. Morimoto, Y. Itoh, K. Yoshimura, M. Kato, and K. Hirota, J. Phys. Soc. Jpn. \textbf{75}, 083709 (2006).

\bibitem{McGuire-17} M. A. McGuire, J. Yan, P. Lampen-Kelley, A. F. May, V. R. Cooper, L. Lindsay, A. Puretzky, L. Liang, S. KC, E. Cakmak, S. Calder, and B. C. Sales, Phys. Rev. Mater. {\bf 1}, 64001 (2017).

\bibitem{Lefransois-16} E. Lefrancois, M. Songvilay, J. Robert, G. Nataf, E. Jordan, L. Chaix, C. V. Colin, P. Lejay, A. Hadj-Azzem, R. Ballou, and V. Simonet, Phys. Rev. B \textbf{94}, 214416 (2016).

\bibitem{Bera-17} A. K. Bera, S. M. Yusuf, A. Kumar, and C. Ritter, Phys. Rev. B \textbf{95}, 094424 (2017).

\bibitem{Lee2012} S. Lee, S. Choi, J. Kim, H. Sim, C. Won, S. Lee, S. A. Kim, N. Hur, and J.G. Park, J. Phys. Condens. Matter \textbf{24}, 456004 (2012).


\bibitem{Korotin2015} D. M. Korotin, V. V. Mazurenko, V. I. Anisimov, and S. V. Streltsov, Phys. Rev. B \textbf{91}, 224405 (2015).

\bibitem{Streltsov2015a} S. Streltsov, I. I. Mazin, and K. Foyevtsova, Phys. Rev. B \textbf{92}, 134408 (2015).
\bibitem{Zivkovic2010} I. Zivkovic, K. Prsa, O. Zaharko, and H. Berger, J. Phys. Condens. Matter \textbf{22}, 56002 (2010).

\bibitem{Kitaev2006} A. Kitaev, Ann. Phys. (N. Y). \textbf{321}, 2 (2006).

\bibitem{Jackeli2009} G. Jackeli and G. Khaliullin, Phys. Rev. Lett. \textbf{102}, 17205 (2009).

\bibitem{banerjee-16} A. Banerjee, C. A. Bridges, J-Q. Yan, A. A. Aczel, L. Li, M. B. Stone, G. E. Granroth, M. D. Lumsden, Y. Yiu, J. Knolle, D.L. Kovrizhin, S. Bhattacharjee, R. Moessner, D. A. Tennant, D. G. Mandrus, and S. E. Nagler, Nature Materials \textbf{15}, 733–740 (2016).

\bibitem{zheng-17} J. Zheng, K. Ran, T. Li, J. Wang, P. Wang, B. Liu, Z-X. Liu, B. Normand, J. Wen, and W. Yu, Phys. Rev. Lett. \textbf{119}, 227208 (2017).

\bibitem{banerjee-2017} A. Banerjee, J. Yan, J. Knolle, C. A. Bridges, M. B. Stone, M. D. Lumsden, D. G. Mandrus, D. A. Tennant, R. Moessner, and S. E. Nagler, Science \textbf{356}, 1055 (2017).


\bibitem{Miura2007} Y. Miura, Y. Yasui, M. Sato, N. Igawa, and K. Kakurai, J. Phys. Soc. Jpn. \textbf{76} 033705 (2007).



\bibitem{Miura-09} Y. Miura, M. Sato, Y. Yamakawa, T. Habaguchi, and Y. Ono, J. Phys. Soc. Jpn. \textbf{78}, 094706 (2009).

\bibitem{Kimber-14} S. A. J. Kimber, I. I. Mazin, J. Shen, H. O. Jeschke, S. V. Streltsov, D. N. Argyriou, R. Valenti, and D. I. Khomskii, Phys. Rev. B \textbf{89}, 081408 (2014).

\bibitem{park} J. Park, T-Y. Tan, D. T. Adroja, A. Daoud-Aladine, S. Choi, D-Y. Cho, S-H. Lee, J. Kim, H. Sim, T. Morioka, H. Nojiri, V. V. Krishnamurthy, P. Manuel, M. R. Lees, S. V. Streltsov, D. I. Khomskii, and J-G. Park, Scientific reports \textbf{6} 25238 (2016).

\bibitem{Na} J. C. Wang, J. Terzic, T. F. Qi, F. Ye, S. J. Yuan, S. Aswartham, S. V. Streltsov, D. I. Khomskii, R. K. Kaul, and G. Cao, Phys. Rev. B \textbf{90}, 161110(R) (2014).

\bibitem{Gapontsev-17} V. V. Gapontsev, E. Z. Kurmaev, C. I. Sathish, S. Yun, J-G. Park, and S. V. Streltsov, J. Phys. Condens. Matter \textbf{29}, 405804 (2017).

\bibitem{VASP} G. Kresse and J. Hafner, Phys. Rev. B \textbf{47},  558, (1993).

\bibitem{V} G. Kresse and J. Furthmuller, Comput. Mater. Sci. \textbf{6} 15, (1996).
\bibitem{PAW} G. Kresse and D. Joubert, Phys. Rev. B \textbf{59} 1758, (1999).

\bibitem{PBE} J. P. Perdew, K. Burke and M. Ernzerhof, Phys. Rev. Lett. \textbf{77}, 3865 (1996).

\bibitem{Andersen-84} O. K. Andersen and O. Jepsen, Phys. Rev. Lett. \textbf{53}, 2571 (1984). 

\bibitem{Segura-2016} M. P. Jimenez-Segura, A. Ikeda, S. Yonezawa, and Y. Maeno, Phys. Rev. B \textbf{93}, 75133 (2016).

\bibitem{Martin-book} R. M. Martin, Electronic Structure: Basic Theory and Practical Methods (Cambridge University Press, 2004).

\bibitem{Streltsov-2016} S. V. Streltsov and D. I. Khomskii, Proc. Natl. Acad. Sci. \textbf{113}, 10491 (2016).

\bibitem{Streltsov-UFN}  S. V. Streltsov and D. I. Khomskii  Physics-Uspekhi \textbf{60}, 1121 (2017).

\bibitem{Anisimov-1991} V. I. Anisimov, J. Zaanen, and O. K. Andersen, Phys. Rev. B \textbf{44}, 943 (1991).

\bibitem{Biroli-2002} G. Biroli and G. Kotliar, Phys. Rev. B \textbf{65}, 155112 (2002).

\bibitem{Hermann-2018} V. Hermann, M. Altmeyer, J. Ebad-Allah, F. Freund, A. Jesche, A. A. Tsirlin, M. Hanfland, P. Gegenwart, I. I. Mazin, D. I. Khomskii, R. Valenti, and C. A. Kuntscher, Phys. Rev. B \textbf{97}, 20104 (2018).



\end{thebibliography}
\end{document}